\documentclass[aps,prl,twocolumn,superscriptaddress,showpacs,preprintnumbers,amsmath,amssymb]{revtex4-1}
\usepackage{times}
\usepackage{color}
\input pdfcolor.tex
\usepackage{graphicx}
\usepackage{colortbl,amsthm,amsmath,amssymb,txfonts}
\usepackage{epsfig,color}
\usepackage[colorlinks,bookmarks=false,citecolor=blue,linkcolor=red,urlcolor=blue]{hyperref}

\begin{document}
 
\title{Thermally induced gaplessness and Fermi arcs in a ``s-wave'' magnetic superconductor}

\author{Madhuparna Karmakar$^{1,2}$ and Pinaki Majumdar} 
\affiliation{
Harish-Chandra Research Institute, HBNI,
Chhatnag Road, Jhunsi, Allahabad 211019, India\\
$^2$ Department of Physics, Indian Institute of Technology, Madras,
Chennai-600036, India.
}
   
\begin{abstract}
An electron system with pre-existing local moments and an effective 
electron-electron attraction can exhibit simultaneous magnetic and 
superconducting order. Increasing the magnetic coupling weakens 
pairing and the ground state loses superconductivity at a critical 
coupling. In the vicinity of the critical coupling magnetic order 
dramatically modifies the quasiparticle dispersion in the 
superconductor, creating low energy spectral weight and significant 
gap anisotropy in the notional `s-wave' state. Using a Monte Carlo 
approach to the Hubbard-Kondo lattice problem we establish a thermal 
phase diagram, for varying magnetic coupling, that corresponds 
qualitatively  to the borocarbide superconductors. In addition to the 
superconducting and magnetic transition temperatures, we identify two 
new thermal scales in this nominal $s$-wave system. These are
associated, respectively, with crossover from gapped to gapless 
superconductivity, and from an anisotropic (nodal) `Fermi surface' 
at low temperature, through a Fermi arc regime, to an isotropic 
Fermi surface at high temperature. Some of the spectral effects are 
already visible in the Ho and Er based borocarbides, others can be 
readily tested.
\end{abstract}

\date{\today}
\maketitle

The interplay of superconductivity 
and magnetism \cite{dagotto2012} is of 
fundamental interest in condensed matter.  
Prominent examples include the high-$T_{c}$ 
cuprates \cite{lee2006}, heavy fermions 
\cite{scalapino2012}, 
and the iron based superconductors \cite{stewart2011}.
The superconductivity in the cuprates
emerges on doping an antiferromagnetic
insulator \cite{lee2006}, in the iron-pnictide  
it emerges from a 
collinear antiferromagnet \cite{stewart2011},
in the iron chalcogenide from bicollinear antiferromagnets
 \cite{fang2008}, while iron selenide superconductors 
are proximate to an antiferromagnetic insulator \cite{guo2010}.
In many of these compounds superconductivity is 
dictated by off-site $d$-wave type pairing
while the magnetic moments arise from electron-electron 
repulsion. Relatively less explored 
is the interplay of $s$-wave pairing and magnetic order
 in, e.g, the rare earth quaternary 
borocarbides (RTBC), a traditional phonon mediated
 superconductor 
\cite{muller2001, gupta1998, michor1999, schneider2009,
 rybalt1999, schultz2011,budko2006}.

The combination of magnetic and superconducting 
ordering tendencies lead
to an unusual electronic state.
Evidence of an unconventional
 superconducting gap in the RTBC’s has
been found in thermal conductivity 
\cite{schneider2009,izawa2002,canfield2001},
and ultrasound attenuation experiments\cite{tagaki2004}. 
Direct evidence was recently 
provided by angle resolved 
photoemission spectroscopy (ARPES) 
\cite{baba2010}, and point contact spectroscopy
\cite{rybalt1999, bobrov2005} 
in YNi$_{2}$B$_{2}$C and LuNi$_{2}$B$_{2}$C - 
suggesting a superconducting gap with point
nodes. HoNi$_{2}$B$_{2}$C shows
gapless superconductivity at finite temperature 
\cite{rybalt1999} while
ErNi$_{2}$B$_{2}$C has a gap
structure which deviates significantly from the
BCS prediction \cite{baba_er}.

Considerable effort has been
 invested in analyzing the ground state of 
these materials 
\cite{machida1980,levin1982,sakai1981,kontani2004,kulic1986,buzdin2005,
paiva2009,fulde2000,mpk2016}, with the 
rare earth de-Gennes (DG) factor, $S(S+1)$, where 
$S$ is the angular momentum, mimicking a
changing magnetic coupling.
These studies suggest the coexistence of
non collinear magnetic order 
with superconductivity.
The superconducting (SC) state becomes 
gapless at a critical magnetic 
coupling $J_{g}$, say, and at a still larger coupling, 
$J_{c} \sim 2J_{g}$ superconductivity is destroyed. 
At  $J > J_{g}$ the
dispersion comprises of as many as eight branches 
(if the non magnetic SC had only two bands),
the density of states exhibit additional van Hove 
singularities, and the low energy spectral
weight maps out a non trivial ``Fermi 
surface'' in these superconductors \cite{mpk2016}.

However, there seems to be little work that addresses the 
simultaneous effect 
of magnetic and superconducting thermal fluctuations
in these superconductors, particularly the
effect of thermally induced magnetic disorder.
The classic Abrikosov-Gorkov (AG) theory \cite{ag}
 describes the impact of random 
uncorrelated moments on the superconductor.
It applies to the regime 
where the moment concentration $\eta$, 
electron-moment coupling $J$, and
pairing gap $\Delta$, satisfy $\eta J^2N(0) \ll \Delta$, 
$N(0)$  being the normal state
density of states at the Fermi level. The theory predicts
 a window of gapless
superconductivity, and finally the loss of SC order, on 
increasing $\eta J^{2}$.
In case of RTBC the moments are on every site, so 
$\eta = 1$, the $J$ and $\Delta$ are 
comparable, and the moments have an
ordered low temperature state. The relatively large
$J$ means that magnetic effects
cannot be treated perturbatively, while spatial
correlation between the moments require a sophisticated
`disorder averaging'.

We use a method that treats the pairing and magnetic effects
on equal footing, and retains the spatial correlation 
between the moments when considering thermal disorder
effects on the electrons.
Our principal findings are
the following. (i)~We map out a thermal phase 
diagram that captures all the qualitative features
of the RTBC family and predict two new thermal
scales: $T_g$ related to gap closure in the superconductor, 
and $T_{an}$ related to 
the appearance of Fermi surface anisotropy.
(ii)~We demonstrate the realization of gapless 
superconductivity, as observed, for parameters 
corresponding to HoNi$_{2}$B$_{2}$C.
(iii)~At moderate magnetic coupling, the scattering from
short range 
magnetic fluctuations leads to a strongly momentum dependent
(non $s$-wave) gap in the superconductor. 
This provides an understanding of the experimental
observations in YNi$_{2}$B$_{2}$C and LuNi$_{2}$B$_{2}$C, which 
reveal an anisotropic SC gap 
in ARPES measurements \cite{baba2010}. 

We study the attractive Hubbard model in
 two dimension on a square lattice in presence of 
Kondo like coupling \cite{mpk2016}:
\begin{equation}
H = H_{0} - \vert  U \vert  \sum_{i}n_{i\uparrow}n_{i\downarrow}  -
J \sum_{i}{\bf S}_{i}.{\bf \sigma}_{i}
\end{equation}
\noindent with, $ H_0 =  \sum_{ij, \sigma}(t_{ij} - \mu \delta_{ij}) 
c_{i\sigma}^{\dagger}c_{j\sigma}$, where $t_{ij}=-t$ for  
nearest neighbor hopping and is zero otherwise. 
${\bf S}_{i}$ is the core spin, arising from $f$ levels, for example, 
in a real material. 
${\bf \sigma}_{i}$ is the electron spin operator. 
 $U$ is the attractive
onsite interaction, giving rise to s-wave
 superconducting order, and $\mu$ is the chemical
potential.
The spin size $S$ dictates the de Gennes factor  
$S(S+1)$ in the rare earths.
We treat the ${\bf S}$ as classical spins, setting 
the size
$\vert {\bf S} \vert =1$, and vary $J$ to mimic 
the varying DG factor. We set $U = 3t$ (although in real materials it is
likely to be smaller) due to system size limitations.

We solve this model by (i)~decoupling  the
Hubbard interaction using an auxiliary
pairing field and retaining only its
static (zero Matsubara frequency) mode, 
(ii)~generating the equilibrium configurations
for the pairing field and local moments via Monte Carlo,
and (iii)~solving the electronic problem in the
`disordered' but spatially correlated backgrounds
by exact diagonalisation.
The method is detailed in the Supplementary materials (SM). 

We use a variety of spatial and spectral indicators to
characterise the phases of the
 system.
These include:
(i)~the 
pairing field structure factor,  $S_{\Delta}({\bf q})$, and 
magnetic structure factor S$_{m}({\bf q})$, 
(ii)~the density of states (DOS), $N(\omega)$ and its
value $N_0$  at the Fermi level, and 
(iii)~the momentum dependent low energy spectral weight
$A{\bf k}, 0$).
The method for calculating these is
 discussed in the SM.

Fig.1 shows the 
the thermal phase diagram 
obtained by our numerical calculations 
and it's comparison with the experimental 
phase diagram of the RTBC family. 
The theory results correspond to 
$U=3t$, and a filling of $n \sim 0.5$.

The ground state at this choice of 
parameters (see SM) has 
magnetic order at the boundary of a spiral
 $(q, \pi)$ and an antiferromagnetic 
$(0, \pi)$ state. In the ground state SC order survives
 upto  $J \sim 0.75t$  and 
comprises of a gapped SC phase for  $0 < J < 0.5t$ followed by
a gapless regime for $ 0.5t < J < 0.75t$.
For $ J > 0.75t$ the ground state is a 
 magnetic metal. 
 
Fig.1(a) shows the thermal phase diagram as
 inferred from experimental 
measurements. There are four major phases:
disordered moment-superconductor (DM-SC),
antiferromagnetic-superconductor (AFM-SC), 
disordered moment-metal 
(DM-M), and antiferromagnetic-metal (AFM-M).
The effective coupling between the core moments
and itinerant electrons increase with spin size $S$,
so the relevant magnetic interaction scale is $\propto
\sqrt{S(S+1)}$. We use this as our $x$ axis, normalising
by the value for Gd.
There are two thermal scales:
the superconducting transition, $T_c$, and 
the magnetic transition, $T_{AF}$.

%-----------------------------------------------------------------------------
\begin{figure}[b]
\centerline{
\includegraphics[height=5cm,width=8.4cm,angle=0]{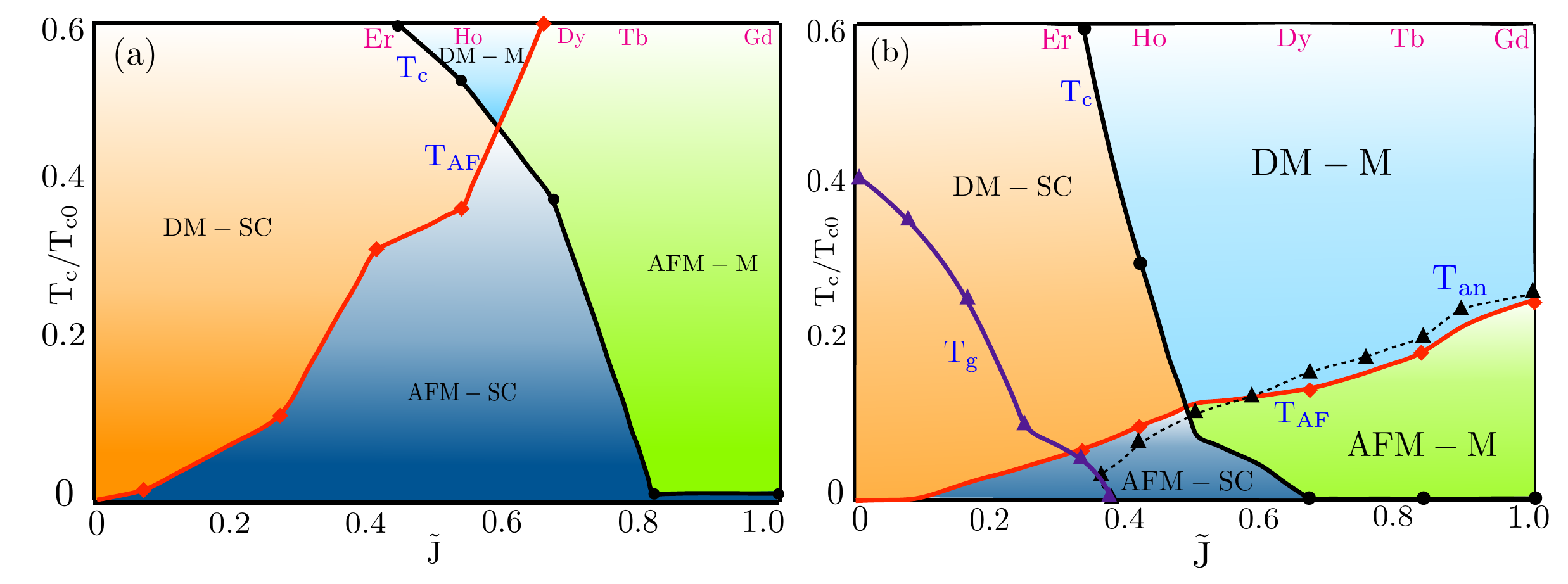}
}
\caption{Color online:  Magnetic coupling-temperature
phase diagram. (a)~Experimental phase diagram of the
RTBC family.  $J^{2} \propto$ the de Gennes (DG) factor
and ${\tilde J}^{2} \sim J^{2}/J_{Gd}^{2}$.
$T_c$ denotes the superconducting transition and
$T_{AF}$ the magnetic transition.
They are normalized by T$_{c0}$, the
superconducting $T_c$ for DG=0.  The phases
are, (i)~disordered moment superconductor (DM-SC),
(ii)~disordered moment metal (DM-M),
(iii)~antiferromagnetic metal (AFM-M) and
(iv)~antiferromagnetic superconductor (AFM-SC).
(b)~Phase diagram from our calculation
at $U=3t$. Thermodynamic phases
are the same as in panel (a).
We show two new temperature scales: $T_g \ll T_c$ where the
spectral gap vanishes in the superconductor, and $T_{an}$
below which the Fermi surface shows significant anisotropy.
}
\end{figure}
%-----------------------------------------------------------------------------

%-----------------------------------------------------------------------
 \begin{figure*}[t]
\begin{center}
\includegraphics[height=5.5cm,width=18cm,angle=0]{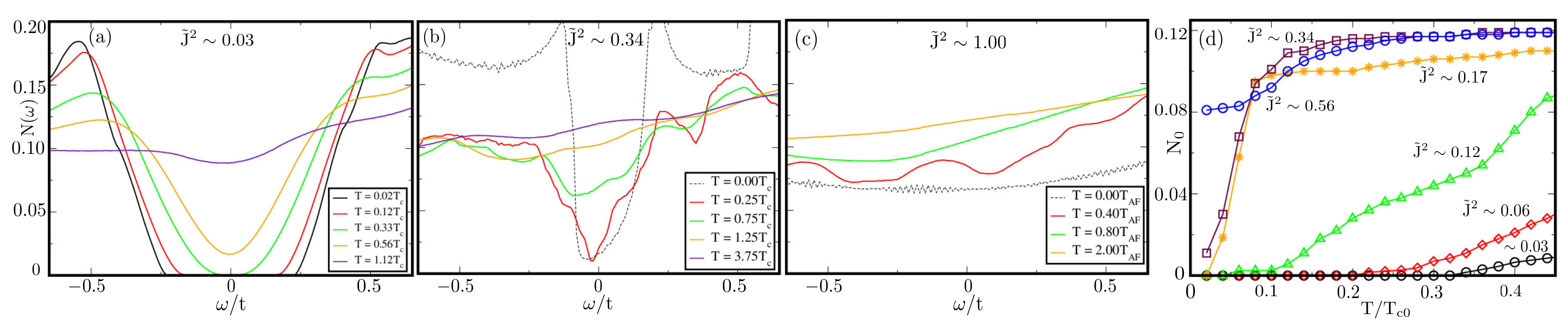}
\end{center}
\caption{Color online: Thermal evolution 
of single particle density of states
(DOS) corresponding to three
different magnetic interaction regimes, viz. (a) weak
($\tilde J^{2} \sim 0.028$), (b) intermediate
($\tilde J^{2} \sim 0.340$) and (c) 
strong ($\tilde J^{2} \sim 1.000$).
Panel (d) shows the DOS at the 
Fermi level (N$_{0}$) plotted as
a function of temperature for 
different magnetic coupling $\tilde J^{2}$.
}
\end{figure*}
%-----------------------------------------------------------------------

With increasing DG factor members of the RTBC family
shows the following behaviour in the ground state: 
(i)~non magnetic 
superconductors (Lu), with no magnetic moments, 
(ii)~(antiferro)magnetic superconductors
 (Tm, Er, Ho and Dy) and 
(iii)~magnetic metal (Tb, Gd). 

In our theory result, Fig.1(b),
there are two main
temperature scales: $T_{c}$ and $T_{AF}$
as in the experiments.  There are,
however,
 two additional thermal scales:  
$T_{g}$ and $T_{an}$,
inferred from the quasiparticle 
spectra. $T_{g}$ indicates
crossover from gapped to gapless superconductivity 
with increasing $T$.
$T_{an}$ marks the onset of pronounced 
momentum dependence of the spectral weight 
at the Fermi level.
This is inferred from the behaviour of the spectral function
$A(${\bf k}, 0$)$, computed from Green's functions in the
finite temperature backgrounds.

Fig.1 indicates that the qualitative features of RTBC physics,
particularly the occurence of the various phases vis-a-vis
experiments,  is reasonably captured by the theory.
To be specific: (i)~in Fig.1(b) the 
AFM-SC phase in the regime  $0 <\tilde J^{2} < 0.6$ 
includes  Tm, Er, Ho and Dy - as is the case with experiments.
(ii)~Tb and Gd (with large magnetic moments) 
are indeed AFM-M, with no SC order.
This correspondence suggests that 
the most suitable members 
to observe {\it gapless} superconductivity are 
Er and Ho. Indeed, a gapless SC 
state at finite $T$ has already been 
reported in HoNi$_{2}$B$_{2}$C
 through point contact
measurements \cite{rybalt1999}. 
In  ErNi$_{2}$B$_{2}$C point contact spectroscopy 
\cite{baba_er} shows clear evidence of the SC gap 
behavior deviating
from the BCS predictions. It was suggested that the 
gap behaviour can be described by a
superconducting theory \cite{machida1980} that incorporates 
magnetic fluctuations. 

%-----------------------------------------------------------------------
\begin{figure*}
\includegraphics[height=12cm,width=16cm,angle=0]{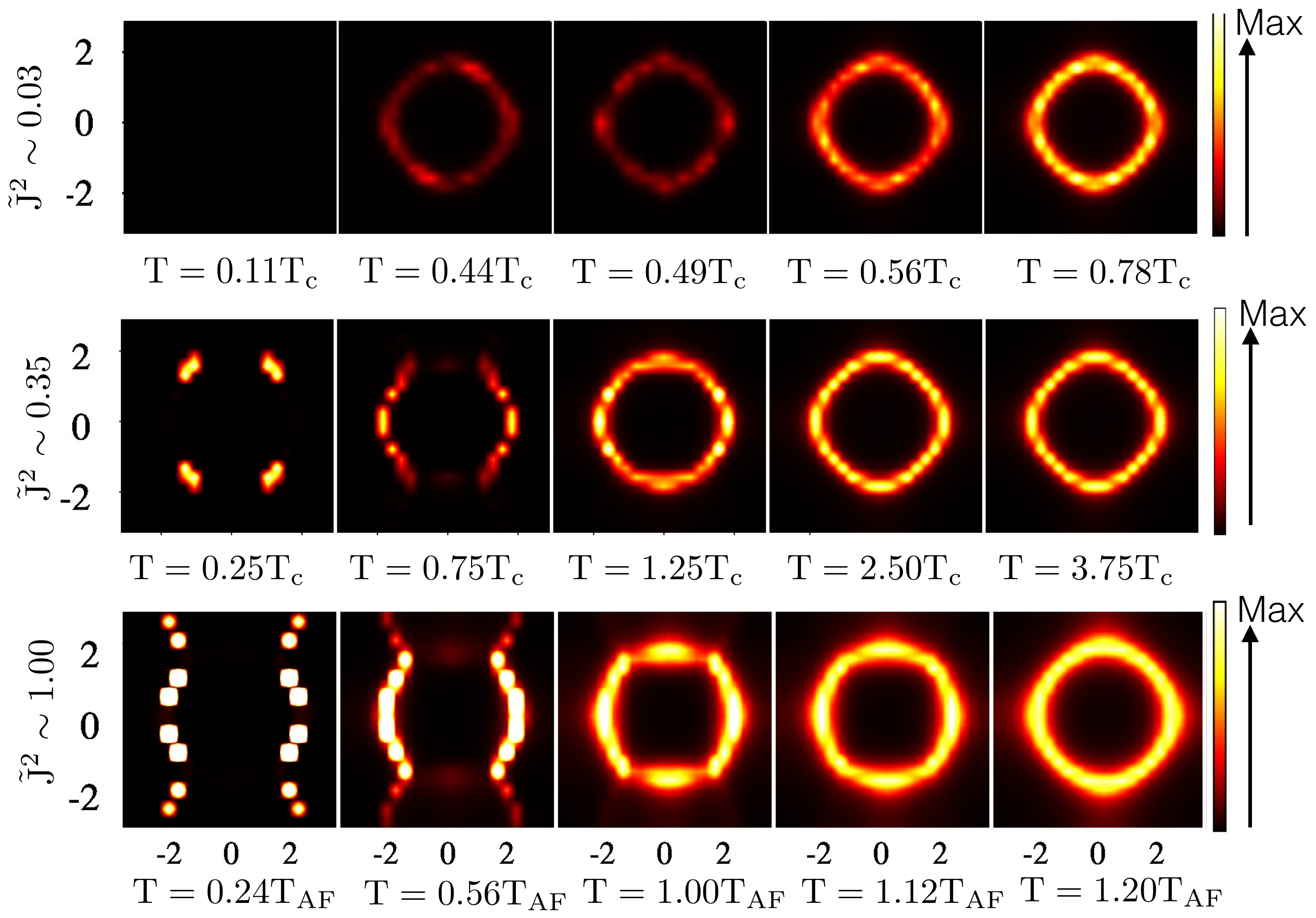}
\caption{Color online: Low energy spectral
weight distribution at $\tilde J^{2} \sim 0.028$,
 $\tilde J^{2} \sim 0.340$ and $\tilde J^{2} \sim 1.000$.
$\tilde J^{2} \sim 0.028$ shows gradual evolution
 from a gapped superconducting state to a pseudogap
through gapless superconducting state before
transiting to a normal (PM-M) state. $\tilde J^{2}\sim
 0.340$ shows accumulation of spectral weight
at isolated points of the ${\bf k}$-space, giving
 rise to an anisotropic gap.
$\tilde J^{2} \sim 1.000$ shows anisotropic Fermi
 surface arising out of band structure effects of
the pure magnetic state.}
\end{figure*}
%-----------------------------------------------------------------------

The effect of magnetic fluctuations 
on RTBC superconductivity
has been demonstrated also through 
inelastic light scattering \cite{lee2000}, 
photoemission spectroscopy 
\cite{oguchi2000}, thermal conductivity 
measurements \cite{canfield2001},
ultrasonic attenuation \cite{tagaki2004},
 ARPES \cite{baba2010} 
experiments on LuNi$_{2}$B$_{2}$C and 
YNi$_{2}$B$_{2}$C which are ``non magnetic'' 
members of the RTBC. While the absence 
of a finite magnetic moment in these 
materials rule out a competing 
magnetic long range order in the ground 
state, short range correlation
among thermally induced `moments' 
is still possible at high 
temperature. 
Our present work does not directly address
 these compounds, since
they involve a `soft magnetic moment' 
arising from Hubbard repulsion,
but the high temperature effects, we 
guess, would be similar to what
we observe here. We will address this separately.

Fig.1(b) however shows 
that our $T_c^0$ scale is too large compared to the 
$T_{AF}$ for Gd,
an obvious inconsistency vis-a-vis the experiments.
This is due to the choice 
$U=3t$, forced by computational constraints. 
In the real RTBC the $U/t \lesssim 1$, but this 
involves large coherence lengths, difficult
to access numerically.
For a closer correspondence with experiments,
in terms of absolute $T_c$ scales, 
the pairing interaction $U$ would have to be smaller.
This would also require the $J$ to be reduced to 
ensure that the magnetism does not suppress
the superconductivity. 
Attaining this within a fully microscopic approach is
difficult and we plan to explore a Ginzburg-Landau scheme
separately.
                 
Fig.2 examines the thermal evolution 
of the single particle DOS at
three magnetic couplings, ${\tilde J}^2$.
At small ${\tilde J}^2$ is similar to that of a  non magnetic 
$s$-wave superconductor. Increasing $T$
drives the gapped low $T$ state to a 
pseudogapped high $T$ state.
At ${\tilde J}^2 \sim 1$ we observe the
 featureless low energy DOS
of the magnetic metal, with gradual
 increase in the DOS with increasing
temperature.
At intermediate ${\tilde J}^2$, where 
$T_{c}$ and $T_{AF}$ are
comparable, we have a 
gapless SC ground state (verified also 
via a Green's function
check \cite{mpk2016}). Rise in 
temperature increases the low energy 
DOS as the system transits from
a AFM-SC to AFM-M and then the DM-M.

In the final panel, Fig.2(d),
 we show the DOS at the Fermi level, $N_{0}$, as a function
of temperature at different magnetic coupling. The weak 
$\tilde J$ regime shows vanishing $N_0$ over a finite $T$ window.
 At intermediate coupling there is 
a finite DOS at the Fermi level, giving rise to gapless SC, and
a prominent $T$ dependence, while at large $\tilde J$ the $N_0$
is large but only weakly $T$ independent.

We next show the emergence of anisotropy in
momentum dependence of the
low energy spectral weight with lowering temperature,
Fig.3, by plotting $A({\bf k},0)$, the spin summed
weight at $\omega=0$. 
Both point contact spectroscopy as 
well as ARPES measurements carried out on rare 
earth borocarbide family gave evidence of 
considerable deviation of the gap structure from
 the naive expectation of a BCS superconductor. 
 The evidence of ``nodal'' 
gap in members of the RTBC family 
has also been inferred from experiments \cite{baba_er}.
A crude connection between the spectral weight to the momentum 
dependent gap is given by $A({\bf k},0) 
\propto e^{- {\Delta({\bf k})/{k_B T}}}$,
where $\Delta({\bf k})$ is the momentum dependent gap.

In Fig.3 the top row shows the SC with weak magnetic
coupling. There is no low energy spectral 
weight at the 
lowest temperature and the weight increases isotropically 
with increasing $T$. This suggests an essentially ${\bf k}$ 
independent gap. Low temperature 
gives rise to a `ghost Fermi surface' which gradually evolves
into a well defined Fermi surface at  $T \sim T_{c}$. 

At intermediate coupling a very unusual gap structure 
emerges out of the competing orders.
The low temperature state at this parameter point is a 
gapless superconductor with spectral weight at 
selected ${\bf k}$-points of the Brillouin zone, decided by 
the magnetic wave vector of the 
underlying order, see the leftmost panel, $0.25T_c$. 
With increasing
temperature, next panel, the pointlike structure has 
broadened into an arc due to magnetic fluctuations. In fact
at $1.25T_c$, where the system is an antiferromagnetic metal,
the Fermi surface, expectedly, is still different from
the tight binding shape. In fact only at $T \sim 2.5T_c$,
when it exits the magnetic phase, does it regain the
the tight binding shape (extreme right of all panels).
This should be visible in the Ho based RTBC, and in 
general in other members of the family which have 
a strongly momentum dependent magnetic susceptibility. 

The bottom row corresponds
to strong coupling where the 
system is a magnetic metal. The Fermi 
surface continues to be anisotropic but
connected. This structure arises due to 
the modified band structure generated by the
magnetic order. 
At T$\sim$0.3T$_{c0}$ the isotropy of the 
Fermi surface is restored as magnetic fluctuations 
ceases to be dominant. 
                                                                               
In conclusion, the present
work is the first theoretical attempt
to describe the thermal behavior of the entire 
borocarbide family within a single framework.
We have mapped out the thermal phase diagram
of this family based on the thermodynamic 
and quasiparticle signatures. Along with the 
transition temperatures,
$T_{c}$ and $T_{AF}$, we identify 
two new thermal scales, 
$T_{g}$ and $T_{an}$, related, respectively,
to the transition from gapped to gapless SC, 
and from 
anisotropic to isotropic SC gap structure.
All this happens within a model where the 
attractive on site interaction tends to
generate an isotropic  $s$-wave gap, while magnetic 
order creates a nodal Fermi surface in the ground
state, and a diffuse Fermi arc like structure at
intermediate temperature below $T_c$. These 
results have a direct bearing on spectroscopy of
the Ho and Er based borocarbides.

\textit{{Acknowledgment}}
We acknowledge use of the High Performance Computing
facility at HRI, Allahabad, India.
 
\bibliographystyle{apsrev4-1}
\bibliography{sc_finite.bib}

\end{document}

\title{Thermally induced gaplessness and Fermi arcs in a ``s-wave'' magnetic superconductor \\ SUPPLEMENTARY MATERIALS}

\author{Madhuparna Karmakar$^{1,2,3}$ and Pinaki Majumdar$^{1,2}$}
\affiliation{$^1$~Harish-Chandra Research Institute, Chhatnag Road, Jhunsi, Allahabad 211019, India.\\
$^2$~Homi Bhabha National Institute, Training School Complex, 
Anushakti Nagar, Mumbai 400085, India.\\
$^3$Department of Physics, Indian institute of technology, Madras, Chennai-600036, India.
}
\renewcommand{\theequation}{S\arabic{equation}}
\renewcommand{\thefigure}{S\arabic{figure}}
\setcounter{figure}{0}
\setcounter{equation}{0}

\maketitle

\section{Model and method}

We study the attractive Hubbard model in two dimension 
on a square lattice in presence of Kondo-like coupling:

\begin{eqnarray}
H & = & H_{0} -\mid U \mid \sum_{i}n_{i\uparrow}n_{i\downarrow} 
- J\sum_{i}{\bf S}_{i}.\sigma_{i}
\end{eqnarray}
\noindent with, $H_{0} = \sum_{\langle ij \rangle,\sigma}(t_{ij}-\mu\delta_{ij})
c_{i\sigma}^{\dagger}c_{j\sigma}$, where $t_{ij}=-t$ for nearest 
neighbor hopping and zero otherwise. ${\bf S}_{i}$ is the core 
spin, arising from the $f$ electrons in real material; $\sigma_{i}$
is the electron spin operator. 

Using a single channel Hubbard-Stratonovich decomposition we 
decompose the four fermion attractive interaction term 
into quadratic fermions in an arbitrary space-time fluctuating 
pairing field. For this we introduce the auxiliary complex 
scalar field $\Delta_{i}(\tau)=\mid \Delta_{i}(\tau)\mid e^{i\theta_{i}(\tau)}$.
A complete treatment of the problem requires retaining 
both space and time dependence of $\Delta$ and can be addressed 
only through Quantum Monte Carlo technique. We however drop 
the $\tau$ dependence and retain the complete spatial dependence 
thereby rendering $\Delta_{i}$ classical. In terms of Matsubara 
frequency we have retained only the $\Omega = 0$ mode. This is 
a fair approximation at high temperature where the energy levels 
are well separated and only the $\Omega = 0$ mode is vital. The 
approximation thus becomes progressively accurate as $T \rightarrow \infty$. 
At $T \neq 0$ the approximation gives fairly accurate results 
and captures the thermal scales correctly. At $T = 0$ the results 
obtained are as good as that from the mean field theory \cite{spa}. 

For the magnetic order, the core spin ${\bf S}_{i}$ is
treated as classical (with a fixed magnitude $S$).
An approximation valid when $2S >> 1$. The angular fluctuations 
of the spin are retained completely. The resulting effective 
Hamiltonian can thus be expressed as:

\begin{eqnarray}
H_{eff} & = & H_{0} + \sum_{i}(\Delta_{i}c_{i\uparrow}^{\dagger}c_{i\downarrow}^{\dagger} + H.c) - J\sum_{i}{\bf S}_{i}.\sigma_{i} \nonumber \\ && 
+ \sum_{i}\frac{\mid \Delta_{i}\mid^2}{U}
\end{eqnarray}    
\noindent where, the last term correspond to the stiffness cost 
associated with the now classical pairing field. The configurations 
$\{\Delta_{i},{\bf S}_{i}\}$ that need to be considered obey 
Boltzmann distribution, obtained by tracing over the electrons:

\begin{eqnarray}
P\{\Delta_{i},{\bf S}_{i}\} \propto Tr_{c,c^{\dagger}}e^{-\beta H_{eff}}
\end{eqnarray} 
\noindent which in turn is related to the free energy of the electrons 
in that configuration. For large and random $\{\Delta_{i}, {\bf S}_{i}\}$
the trace has to be taken numerically. For the finite temperature 
results presented in this paper we generate equilibrium configurations 
by using Metropolis algorithm for $\{\Delta_{i},{\bf S}_{i}\}$ and 
estimate the update cost by diagonalizing the electron Hamiltonian 
$H_{eff}$ for every microscopic move. This numerically expensive method 
has been rendered tractable by applying travelling cluster approximation 
(TCA) wherein instead of diagonalizing the entire lattice for each attempted 
update a smaller cluster surrounding the update site is diagonalized 
\cite{tca}.  

\subsection{Indicators}

Using the equilibrium configurations obtained through simulated annealing
we calculate the thermodynamic and 
quasiparticle indicators required to characterize the phases,  
viz. the pairing field structure factor ($S_{\Delta}({\bf q})$), magnetic structure
factor ($S_{m}({\bf q})$), low energy weight distribution ($A({\bf k}, 0)$) and the electronic density of states
($N(\omega) = \sum_{\bf k}A({\bf k}, \omega)$) (where $A({\bf k},\omega)$
is the momentum resolved spectral function) defined as,

\begin{eqnarray}
S_{\Delta}({\bf k}) & = & (1/N)\sum_{ij}\langle \Delta_{i}.\Delta_{j}^{*}\rangle 
e^{i {\bf q.r}} \\ S_{m}({\bf k}) & = & (1/N)\sum_{ij}\langle {\bf S}_{i}.{\bf S}_{j}\rangle
e^{i{\bf q.r}}\\ A({\bf k}, \omega) & = & \sum_{\sigma}(\mid u_{n, \sigma}^{\bf k}\mid^{2}\delta(\omega - E_{n}) \\ \nonumber &&
+ \mid v_{n, \sigma}^{\bf k}\mid^{2}\delta(\omega + E_{n}))
\end{eqnarray}
here, ${\bf S}_{i} = (S_{i}^{x}\hat i + S_{i}^{y}\hat j + S_{i}^{z}\hat k)$, where,
\begin{eqnarray}
S_{i}^{x} & = & S\sin\alpha_{i}\cos\phi_{i} \\ S_{i}^{y}  & = & S\sin\alpha_{i}\sin\phi_{i} \\
S_{i}^{z} & = & S\cos\alpha_{i}
\end{eqnarray}
\noindent where, $\alpha_{i}$ and $\phi_{i}$ are the polar and azimuthal angles 
 of the magnetic core spin, respectively.
$u_{n, \sigma}^{\bf k}$ and $v_{n, \sigma}^{\bf k}$ are the 
Bogoluibov de Gennes (BdG) eigen vectors with the corresponding eigen value $E_{n}$.

\section{Ground state phase diagram at $U=3t$}

To determine the ground state we use a variational scheme. 
We minimize the energy over a restricted family of 
$\{\Delta_{i}, {\bf S}_{i}\}$ configurations. We assume 
$\Delta_{i} = \Delta_{0}$, a site independent real quantity
and for the magnetic order we assume spiral  
configurations where the polar angle $\alpha_{i} = \pi/2$ 
and the azimuthal angle $\phi_{i}$ is periodic, with 
$S_{i}^{z}=0$, $S_{i}^{x}$=cos$({\bf q. r_{i}})$ and $S_{i}^{y}$ = sin$({\bf q. r_{i}})$.
The allowed wave vectors $\{q_{x}, q_{y}\}$ are of the form $2\pi n/L$, where 
n = 1, 2, 3, .... We minimize the energy over $\{q_{x}, q_{y}\}$ and $\Delta_{0}$
for a fixed $\mu$, J and U. We obtain the optimized $\{\Delta_{0}, {\bf q}\}$ 
configuration for a fixed $\mu$ and then evaluate the density n from it so as 
to obtain the phase diagram in the n-J space for any given U/t \cite{mpk2016}.

Fig.S1 shows the ground state n-J phase diagram at U=3t. In the absence of any
superconducting order the ground state is characterized only 
by the magnetic ordering vector ${\bf Q}$ (where ${\bf Q}$ is the optimized $\{q_{x}, q_{y}\}$).
The small J/t
limit is governed by RKKY interaction with the order 
being dictated by the maxima of spin susceptibility. The magnetic 
state depends on $\mu$ or filling n. At low filling the magnetic 
state correspond to ferromagnetic order with ${\bf Q} = \{0, 0\}$.
With increasing filling the magnetic state undergoes transition to 
a $\{0, q\}$ state at intermediate filling, followed by an antiferromagnetic 
$\{0, \pi\}$ state to a spiral $\{q, \pi\}$ state and finally to 
a Neel antiferromagnet $\{\pi, \pi\}$ at half filling n = 1.

%----------------------------------------------------------------
\begin{figure}
\begin{center}
\includegraphics[height=6cm,width=7cm,angle=0]{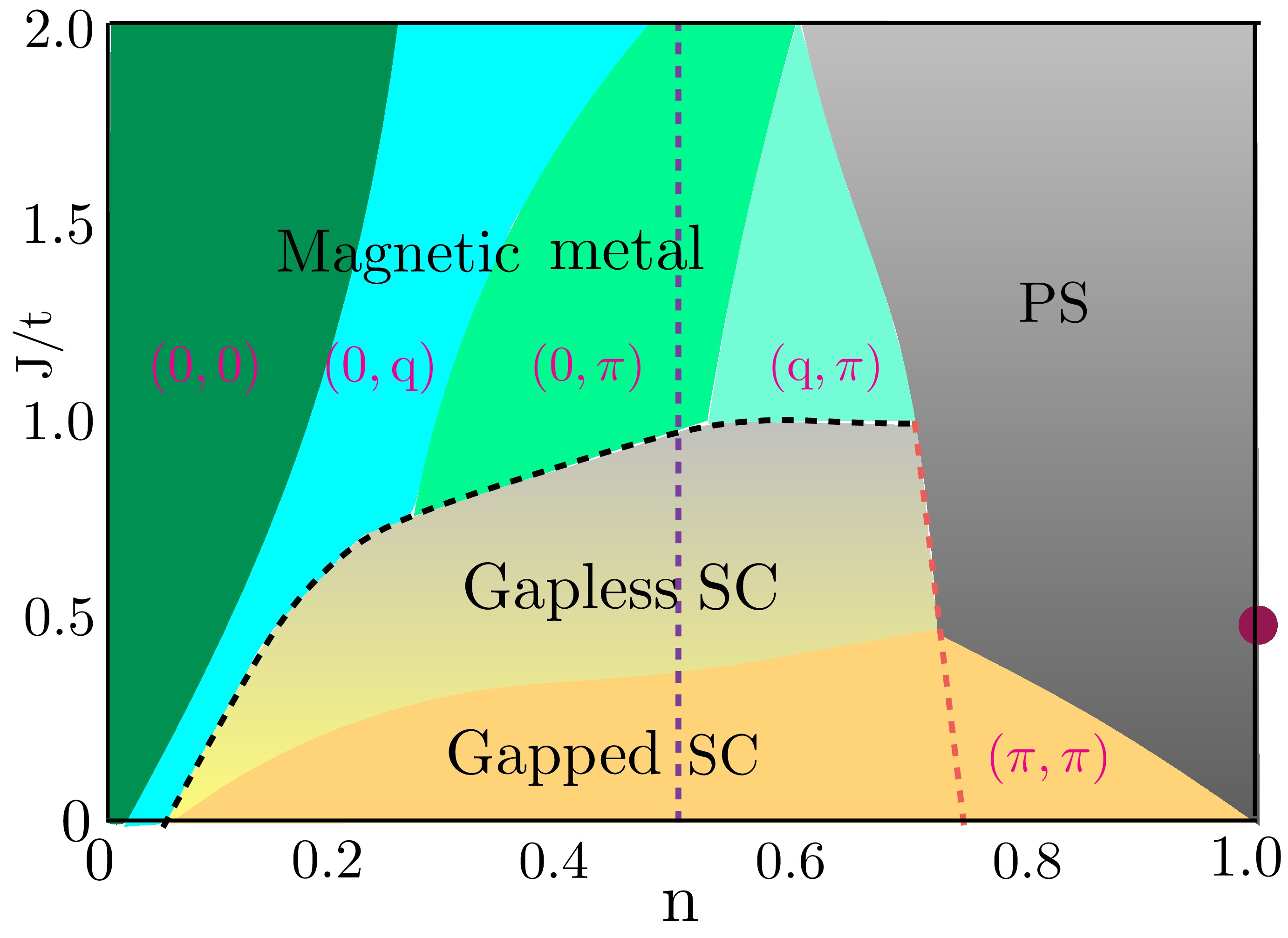}
\end{center}
\caption{Color online: Ground state n-J phase diagram at U=3t
  showing the phases (a) gapped superconductor, (b) gapless superconductor,
  (c) magnetic metal and (d) phase separation. The black dashed line correspond to 
J$_{c}$ (see text). The vertical dotted line represents the cross section at which 
the thermal data is presented in the main text.}
\end{figure}
%----------------------------------------------------------------

At U = 3t and J = 0 there is the usual ${\bf k}\uparrow$ and ${\bf -k}\downarrow$ pairing.
At a small finite J/t the superconducting state
is weakly modified by the magnetic order. The pairing field 
($\Delta_{0}$) undergoes 
suppression but superconducting state continues to be gapped upto a 
coupling, J$_{g}$, say. The maximum J$_{g}$ $\sim$ 0.3t is obtained 
for a filling of n $\sim$ 0.75. For J $>$ J$_{g}$ superconducting state 
is significantly modified by the magnetic order. Along with the 
suppression in $\Delta_{0}$ there is now emergence of subgap features 
in the DOS at the Fermi level. With a finite DOS at the Fermi level 
superconductivity is rendered gapless over the regime 
J$_{g}$ $<$ J $<$ J$_{c}$, where,  J$_{c}$ is the critical coupling beyond 
which superconductivity gives way to magnetic metal.
J$_{c}$/t increases with increasing magnetic coupling and has 
its maxima of J$_{c}$ $\sim$ 0.8t for the filling of n $\sim$ 0.7.
In the limit 
of large filling n $\approx$ 1 there is phase separation regime over 
most part except for the small J/t regime, 
where a gapped superconducting state survives in spite of an underlying 
$\{\pi, \pi\}$ order. At and near half filling the DOS is always gapped 
at the Fermi level. Upto J $\sim$ 0.3t there exists a superconducting 
gap at the Fermi level. For J $>$ 0.3t the gap arises out of the 
$\{\pi, \pi\}$ magnetic order.      

%--------------------------------------------------------------------
\begin{figure}
\begin{center}
\includegraphics[height=5cm,width=8cm,angle=0]{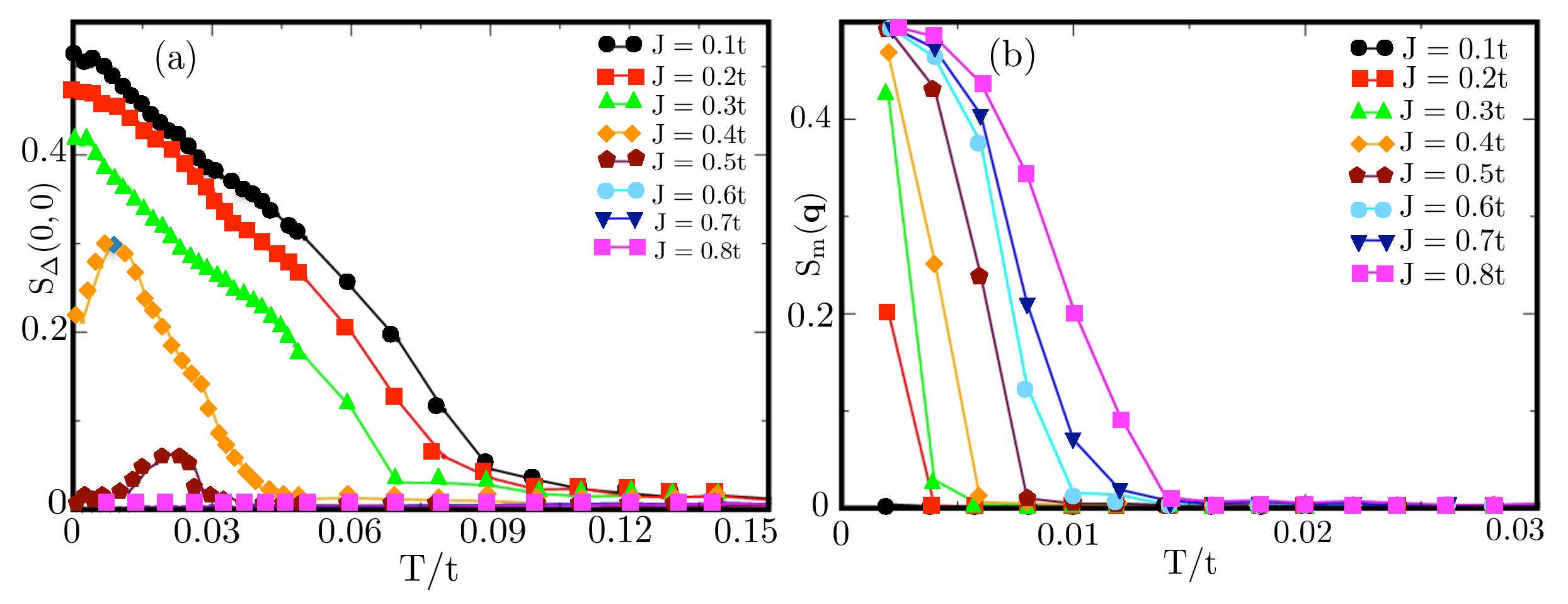}
\end{center}
\caption{Color online: Thermal evolution of structure factor peak for
 (a) pairing field ($S_{\Delta}(0, 0)$) and (b) magnetic order ($S_{m}({\bf q})$)
 at $U=3t$ and different magnetic coupling $J/t$.}
\end{figure}  
%--------------------------------------------------------------------

The thermal evolution of the ground state phases were tracked 
using the superconducting and magnetic structure factors.  
We plot the peak amplitude of superconducting (S$_{\Delta}$(0)) and 
magnetic (S$_{m}$(${\bf q}$)) structure factor in Fig.S2, at different 
magnetic couplings. 
The superconducting order gets progressively suppressed 
with increasing J/t. At the density regime we are in 
a non collinear (${\bf q}, \pi$) magnetic order is realized at all J/t. The 
magnitude of ${\bf q}$ shows only weak dependence on J/t. 

\section{Additional issues}
  
\subsection{ The general validity of the approximations used here}
  
The results presented in this paper are based on the attractive Hubbard 
model with Kondo like coupling. The interaction is decomposed in terms of
auxiliary complex scalar field $\Delta_{i}(\tau) 
= \mid \Delta_{i}(\tau)\mid e^{i\theta_{i}(\tau)}$ using  Hubbard-Stratonovich
decomposition which converts the four fermion term into quadratic fermions
in an arbitrary space-time fluctuating pairing field. For the magnetic order,
we have a quantum ``spin $S$'' with magnetic moment ${\bf S}_{i}$ coupled to the
electrons. There are two main approximations involved in this calculation. 
We discuss them pointwise. (a) In order to make the problem numerically 
tractable we have dropped the $\tau$ dependence of the pairing field thereby treating
it as classical. The spatial fluctuations of $\Delta_{i}$ are retained which are
essential to capture the finite temperature behavior. The approximation gets progressively
accurate as T $\rightarrow \infty$, is fairly accurate at T $\neq$ 0 and akin to mean field
theory at T=0. Since the present paper discusses about the finite temperature
behavior of the system, this is a reasonable approximation and captures the thermal
scales successfully. (b) ${\bf S}_{i}$ has been treated as classical spin
but its angular fluctuations are retained at finite temperature. This approximation 
is valid when 2S $\gg$ 1. In case of the family of rare earth borocarbides (RTBC), the 
4f shells for the magnetic superconductor involves 2S $\sim$ 3-5 thereby making
the ``classical'' spin approximation reasonable. However, the behavior of low moment and 
nonmagnetic superconductors such as Tm and Lu can not be suitably captured 
with our present approximation.
            
\subsection{Survival of the effects to U/t $\ll$ 1, and the various T scales that one expects then}

Owing to the numerical complexities it is difficult to study the thermal physics beyond a lattice
size of 30$\times$30 within a reasonable computation time. The results presented in this paper 
corresponds to a typical interaction of U = 3t. We consider it as a representative of the 
``weak'' coupling regime. The real materials (RTBC) however has  U $\ll$ 1t and thus a 
suppressed pairing field. Consequently, both the superconducting gap and the thermal scales 
are strongly suppressed. Superconductivity in this case would be realized over a narrow window 
in the J-T phase diagram. However, it could still be classified into gapped and
gapless regimes. There is a small but nonzero J$_{g}$ and within the regime
J$_{g}$ $<$ J $<$ J$_{c}$ there would be finite DOS at the Fermi level. We emphasize that 
the emergence of gapless superconductivity is not an artifact of strong interaction
and is expected to be observable in the real materials, albeit in a narrow parameter
window. 
                              
\subsection{ Possible Landau-Ginzburg functional}

In order to develop an insight on the system under consideration through 
a phenomenological Landau-Ginzburg (LG) theory we carry out a systematic 
expansion of the magnetic and superconducting order parameters. The 
resulting free energy functional takes the form,
\begin{eqnarray}
F_{eff} & = & F_{\Delta} + F_{J} + F_{\Delta, J}, \nonumber \\ 
F_{\Delta} &= & \sum_{ij}a_{ij}\Delta_{i}\Delta_{j}^{*} + \sum_{ijkl}b_{ijkl}\Delta_{i}\Delta_{j}^{*}\Delta_{k}\Delta_{l}^{*} + O(\Delta^{6}), \nonumber \\ F_{J} & = & \sum_{ij}J_{ij}^{(2)}{\bf S}_{i}.{\bf S}_{j} + \sum_{ijkl}J_{ijkl}^{(4)}({\bf S}_{i}.{\bf S}_{j}{\bf S}_{k}.{\bf S}_{j} + ...) + ..., \nonumber \\ F_{\Delta, J} & = & \sum_{ijkl}[c_{ijkl}\Delta_{i}\Delta_{j}^{*}{\bf S}_{k}.{\bf S}_{l} + H.c]+ ...,
\end{eqnarray}
where, $a_{ij} \sim -\chi_{ij}^{P} + (1/U)\delta_{ij}$, $\chi_{ij}^{P}$ being the nonlocal pairing 
susceptibility of the free Fermi system, and $b_{ijkl}$ arises from a convolution of four free Fermi
Green's functions. $J_{ij}^{(2)} \sim -J^{2}\chi_{ij}^{S}$, where $\chi_{ij}^{S}$ is the nonlocal 
spin susceptibility of the free-electron system, leading to the RKKY interaction and $J^{(4)}$, like 
$b_{ijkl}$, involves a four Fermi cumulant. $c_{ijkl}$ can be constructed again from a combination of four Green's functions \cite{mpk2016}.

The terms above define a relatively low order classical field theory on a lattice. $H_{\Delta}$ 
involves the first two terms in the superconducting LG theory, and $H_{J}$ describes the leading 
interaction coupling magnetic moments. $H_{\Delta, J}$ indicates how the two orders modify each 
other. All of this holds when $\Delta_{i}$ and J${\bf S}_{i}$ are $\lesssim$ t.